\documentclass[aps,prb,twocolumn,superscriptaddress]{revtex4-1}
\usepackage{amssymb}
\usepackage{amsmath}
\usepackage{graphicx}
\usepackage{epsfig}
\usepackage{color}

\begin{document}

\title{Dynamical signature of moire pattern in non-Hermitian ladder}
\author{X. M. Yang}
\affiliation{School of Physics, Nankai University, Tianjin 300071, China}
\author{X. Z. Zhang}
\email{zhangxz@tjnu.edu.cn}
\affiliation{College of Physics and Materials Science, Tianjin Normal
University, Tianjin 300387, China}
\author{C. Li}
\affiliation{School of Physics, Nankai University, Tianjin 300071, China}
\author{Z. Song}
\email{songtc@nankai.edu.cn}
\affiliation{School of Physics, Nankai University, Tianjin 300071, China}

\begin{abstract}
We study the dynamical behavior of a non-Hermitian moire superlattice
system, which consists of two-coupled SSH chains with staggered imaginary
on-site potentials. There are two main spatial regions, in which systems are
in unbroken symmetric phases with fully real spectrum, appearing
periodically along the ladder. We show that the two quantum phases are
dimerized and tetramerized, which determine the distinct dynamical
behaviors. Dirac probability can oscillate periodically, increase
quadratically and increase exponentially, which correspond to the unbroken
phase, exceptional point and the broken phase of the tetramerized region. In
comparison, the Dirac probability can exhibit high-frequency oscillation in
the dimerized region. These phenomena demonstrate the dynamical signature
and provide insightful information of the moire pattern in the non-Hermitian
regime.
\end{abstract}

\maketitle

\section{Introduction}

\label{sec_intro}

One of the unique features of a non-Hermitian system is the violation of
conservation law of the Dirac probability, based on which, the complex
potential is employed to describe open systems phenomenologically \cite{Muga}%
. Furthermore, unconventional propagation of light associated with the
gain/loss has been demonstrated by engineering effective non-Hermitian
Hamiltonians in optical systems \cite{Guo,K. G. Makris,Z. H. Musslimani,S.
Klaiman,S. Longhi,Ganainy,Zheng,Graefe}. Around exceptional point (EP), many
unique optical phenomena have been observed, ranging from loss-induced
transparency \cite{Guo}, power oscillations violating left-right symmetry
\cite{Ruter}, low-power optical diodes \cite{PengB}, to single-mode laser
\cite{FengL,Hodaei}. A fascinating phenomenon of non-Hermitian optical
systems in the application aspect is the gain-induced detection, such as
enhanced spontaneous emission \cite{LinZ}, enhanced nano-particle sensing
\cite{Wiersig} as well as the amplified transmission in the optomenchanical
system \cite{LiuYL,ZhangXZ}. Both theoretical and experimental works not
only give an insight into the dynamical property of the non-Hermitian
Hamiltonian but also provide a platform to implement the novel optical
phenomenon.

Recently there has been a growing interest in the influence of the moire
pattern in physical systems. The moire pattern as a new way to apply
periodic potentials in van der Waals heterostructures to tune electronic
properties, has been extensively studied \cite%
{Ponomarenk,Dean,Hunt,Gorbachev,Song,Jung}. Many interesting phenomena have
been observed in the heterostructure\ materials with small twist angles and
mismatched lattice constants. Moire patterns in condensed matter systems are
produced by the difference in lattice constants or orientation of two 2D
lattices when they are stacked into a two-layer structure. The aim of this
paper is to demonstrate the phenomenon of the moire pattern in a
non-Hermitian system via a dynamical process. A fascinating feature of a
non-Hermitian system is the existence of exceptional points, at which two
eigenstates coalesce \cite{Keldysh,Kato,Moiseyev,Berry2004,Heiss2012}. The
dynamics of the system with parameters far away from, near and at the EP,
exhibits extremely different behaviors \cite{ZhangXZ2013,WangP}. (i) When
the system is far from or near EP but with a finite energy gap $\epsilon $,
the dynamics is a periodic oscillation with associated Dirac probability
oscillating in the period of time $2\pi /\epsilon $. (ii) When the system is
at EP, the Dirac probability increases quadratically with time. (iii) When
the system has complex levels, the Dirac probability increases exponentially
with time. The rich variety of dynamical behaviors can show up periodically
along the ladder.

In the paper, we study a modified non-Hermitian ladder system, which
consists of two SSH chains with staggered imaginary potentials. The
irregular structure arises from the slight difference of lattice constants
between two legs. There are three types of approximate regular ladder
structures, with different phases. These phenomena indicate that these three
regions should have distinguishable dynamical behaviors, which are the
signature of moire patterns. The reality of the spectrum is sensitive to the
symmetry of the structure. The corresponding non-Hermiticity enhances the
influence of the effect of moire patterns, that are apparent in the dynamics
of the non-Hermitian system. We show that the dynamics is profoundly changed
by slightly mismatched lattice constants associated with long period moire
patterns.

This paper is organized as follows. In Section \ref{model}, we present the
model Hamiltonian and analyze the structure of the lattice. In Sections \ref%
{Tetramerized phase} and \ref{Dimerized phase}, we investigate\ the quantum
phase diagrams of two typical lattices based on the exact solutions,
respectively. Section \ref{Dynamical signatures} introduces the simple
version of two types of lattice, which capture the main original dynamical
behaviors. \ref{Moire pattern} devotes to the numerical simulation of the
model, revealing the dynamical signature of moire patterns.\ Finally, we
give a summary and discussion in Section \ref{summary}.

\begin{figure*}[tbp]
\centering
\includegraphics[height=12cm,width=18cm]{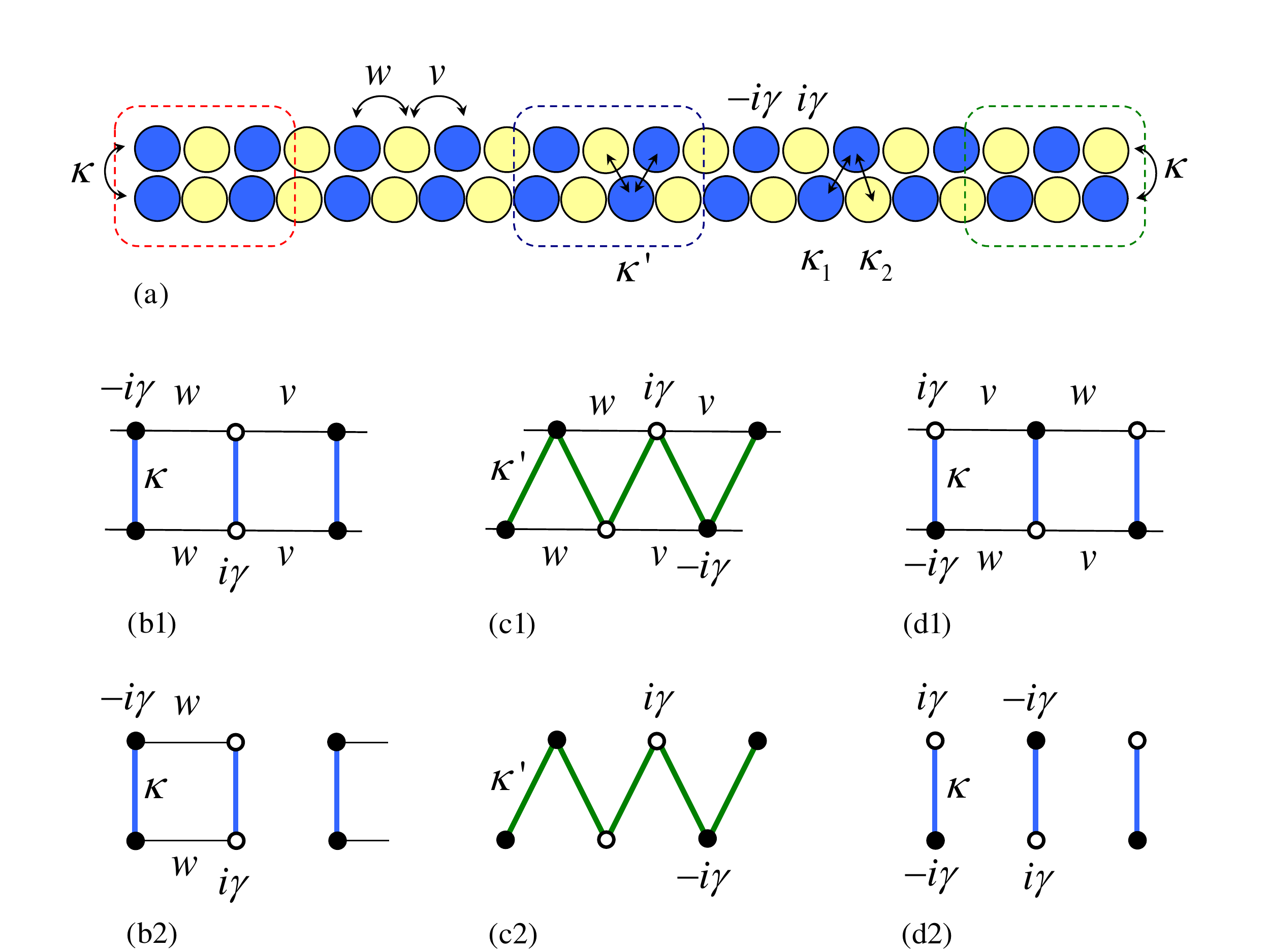}
\caption{(Color online) Schematic illustration of the modified non-Hermitian
two-leg ladder system. It consists of two SSH chains with staggered
imaginary potentials. The irregular structure arises from the slight
difference of lattice constants between two legs. There are three types of
approximate regular ladder structures (circled by the red, blue and green
dotted lines, respectively), which appear periodically in a large scale. The
inter-leg hopping rates are $\protect\kappa $, $\protect\kappa ^{\prime }$, $%
\protect\kappa _{1}$, and $\protect\kappa _{2}$, in various regions,
respectively. (b1), (c1) and (d1) are drafts of three typical structures in
different regions. (b2), (c2) and (d2) are drafts of the reductions of the
structures in (b1), (c1) and (d1), respectively, in the limit case $\protect%
\kappa $, $\protect\kappa ^{\prime }\gg w\gg \protect\nu $ limit. (b2-d2)
capture the main features of (b1-d1)\textbf{. }Systems (b2) and (d2) have
fully real spectra for small enough $\protect\gamma $, but distinguishable
dynamical behaviors (see text). (c2) can also have full real spectrum when $%
\protect\gamma $ is small enough, which can be seen from Fig. 3(b). These
phenomena indicate that three regions should have distinguishable dynamical
behaviors, which are the signature of moire pattern.}
\label{fig1}
\end{figure*}

\section{Model}

\label{model}

In material science, moire patterns are usually produced by stacking two
two-dimensional (2D) crystals into van der Waals heterostructures with an
twist angle. As a non-Hermitian variant of the moire pattern, we take a
simple example by stacking two one-dimensional chains. We consider a two-leg
ladder system with the Hamiltonian

\begin{equation}
H=H_{1}+H_{2}+H_{12}\mathrm{,}
\end{equation}%
where $H_{\lambda }$\ ($\lambda =1,2$)\ describes the independent SSH chain
with staggered imaginary on-site potentials%
\begin{eqnarray}
H_{\lambda } &=&\sum\limits_{l}\left( w\left\vert 2l-1,\lambda \right\rangle
\left\langle 2l,\lambda \right\vert +v\left\vert 2l,\lambda \right\rangle
\left\langle 2l+1,\lambda \right\vert \right)   \notag \\
&&+\mathrm{H.c.}+i\gamma \sum\limits_{l}(-1)^{l}\left\vert l,\lambda
\right\rangle \left\langle l,\lambda \right\vert \mathrm{,}
\end{eqnarray}%
and $H_{12}$\ is the inter-chain tunneling term%
\begin{equation}
H_{12}=\sum\limits_{l^{\prime },l}\kappa _{ll^{\prime }}\left\vert
l,1\right\rangle \left\langle l^{\prime },2\right\vert +\mathrm{H.c..}
\end{equation}%
Here basis $\left\{ \left\vert l_{1},1\right\rangle ,\left\vert
l_{2},2\right\rangle ,l_{\lambda }\in \left[ 1,2N_{\lambda }\right] \right\}
\ $is an orthonormal complete set, satisfying $\langle l,\lambda \left\vert
l^{\prime },\lambda ^{\prime }\right\rangle =\delta _{ll^{\prime }}\delta
_{\lambda \lambda ^{\prime }}$. Unlike the usual case of a ladder, $N_{1}$
and $N_{2}$ are not identical in the present work. The inter-leg tunneling
amplitude $\kappa _{ll^{\prime }}$\ depends on positions of $\left\vert
l,1\right\rangle $ and $\left\vert l^{\prime },2\right\rangle $. The
geometry of the modified two-leg ladder is illustrated in Fig. 1. There are
three typical cases: (i) a site on a leg only couples to a single site with
the same imaginary potential on another leg; (ii) a site on a leg only
couples to a single site with the opposite imaginary potential on another
leg; (iii) a site on a leg couples to two sites on another leg.

We consider the case that lattice constants of two chains are slightly
different. In a certain region, the structure of the ladder can be regarded
as uniform in a large scale. The local dynamics obeys the corresponding
uniform Hamiltonian.

\section{Tetramerized phase}

\label{Tetramerized phase}

We consider the first typical regular ladder system which is illustrated in
Fig. 2(a). The Hamiltonian has the form%
\begin{eqnarray}
&&H_{\mathrm{T}}=\sum\limits_{l=1}^{N}\sum\limits_{\lambda =1,2}\left(
w\left\vert 2l-1,\lambda \right\rangle \left\langle 2l,\lambda \right\vert
+v\left\vert 2l,\lambda \right\rangle \left\langle 2l+1,\lambda \right\vert
\right)  \notag \\
&&+\kappa \sum\limits_{l=1}^{2N}\left\vert l,1\right\rangle \left\langle
l,2\right\vert +\mathrm{H.c.}  \notag \\
&&+i\gamma \sum\limits_{l=1}^{2N}\sum\limits_{\lambda
=1,2}(-1)^{l}\left\vert l,\lambda \right\rangle \left\langle l,\lambda
\right\vert \mathrm{,}
\end{eqnarray}%
where the boundary condition is $\left\vert 2N+1,\lambda \right\rangle
\equiv \left\vert 1,\lambda \right\rangle $. As illustrated in Fig. 2(a), it
satisfies the $\mathcal{PT}$-symmetry. Here, the time reversal operation $%
\mathcal{T}$ is such that $\mathcal{T}i\mathcal{T}=-i$, while the effect of
the parity is such that $\mathcal{P}\left\vert l,\lambda \right\rangle
=\left\vert 2N-l+1,\lambda \right\rangle $ for $\lambda =1,2$. Applying
operators $\mathcal{P}$ and $\mathcal{T}$ on the Hamiltonian $H_{\mathrm{T}}$%
, one has $\left[ \mathcal{T},H_{\mathrm{T}}\right] \neq 0$ and $\left[
\mathcal{P},H_{\mathrm{T}}\right] \neq 0$, but $\left[ \mathcal{PT},H_{%
\mathrm{T}}\right] =0 $. According to the non-Hermitian quantum theory, such
a Hamiltonian may have fully real spectrum within a certain parameter
region. The boundary of the region is the critical point of quantum phase
transition associated with the $\mathcal{PT}$-symmetry breaking. In the
following, we will diagonalize this Hamiltonian and get the phase diagram.

We note that such a system also has another symmetry under the exchange of
two chains, i.e., $\left\vert l,1\right\rangle \leftrightarrows \left\vert
l,2\right\rangle $. This symmetry ensures the conservation of bonding or
antibonding state between two sites coupled by $\kappa $. We refer the
collective bonding or antibonding state as to dimerized phase. Taking the
linear transformation

\begin{equation}
\left\vert l,\sigma \right\rangle =\frac{1}{\sqrt{2}}\left( \left\vert
l,1\right\rangle +\sigma \left\vert l,2\right\rangle \right) ,
\end{equation}%
with $\sigma =\pm $, the Hamiltonian can be rewritten as%
\begin{eqnarray}
&&H_{\mathrm{T}}=H_{+}+H_{-}\mathrm{,} \\
&&H_{\sigma }=\sum\limits_{l=1}^{N}\left( w\left\vert 2l-1,\sigma
\right\rangle \left\langle 2l,\sigma \right\vert +v\left\vert 2l,\sigma
\right\rangle \left\langle 2l+1,\sigma \right\vert \right)  \notag \\
&&+\mathrm{H.c.}+i\gamma \sum\limits_{l=1}^{2N}(-1)^{l}\left\vert l,\sigma
\right\rangle \left\langle l,\sigma \right\vert +\kappa \sigma
\sum\limits_{l=1}^{2N}\left\vert l,\sigma \right\rangle \left\langle
l,\sigma \right\vert \mathrm{.}
\end{eqnarray}%
Sub-Hamiltonian $H_{\sigma }$ satisfies $\left[ H_{+},H_{-}\right] =0$,
representing two independent non-Hermitian SSH chains but\ with opposite
chemical potentials $\pm \kappa $, which has been studied in the previous
work \cite{HWH}. It turns out that the spectrum $\epsilon _{k}^{\sigma }$
for a single chain $H_{\sigma }$\ is

\begin{equation}
\epsilon _{k}^{\sigma }=\sigma \kappa \pm \sqrt{\left( \epsilon
_{k}^{0}\right) ^{2}-\gamma ^{2}},  \label{spectrum}
\end{equation}%
with%
\begin{equation}
\epsilon _{k}^{0}=\sqrt{4wv\cos ^{2}\left( \frac{k}{2}\right) +\left(
w-v\right) ^{2}},
\end{equation}%
which consists of two branches separated by an energy gap
\begin{equation}
\Delta =\sqrt{\left( w-v\right) ^{2}-\gamma ^{2}}.
\end{equation}%
Obviously, it displays a full real spectrum within the region of $\left(
w-v\right) ^{2}\geq \gamma ^{2}$. Beyond this region, the imaginary
eigenvalue appears and the $\mathcal{PT}$ symmetry of the corresponding
eigenfunction is broken simultaneously according to the non-Hermitian
quantum theory. The phase diagram is plotted in Fig. \ref{fig3}(a).\ We note
that the dimerization along the legs still exists ($w\neq v$), when the gap
vanishes in such a non-Hermitian model. In the case of $\kappa \gg w\gg \nu $%
, the combination of two types of dimerizations, inter- and intra-leg
dimers, result in tetramers. We refer the collective tetramerized states as\
to tetramerized phase. The extremely tetramerized phase is characterized by
the ground states of the system with $\nu =0$ (see Fig. \ref{fig1}(b2)).
%%%%%%%%%%%%%%%%%%%%%%%%%%%%%%%%%%%%%%%%%%%%%%%%%%%%%%%%%%%%%%%%%%%%%%%

\begin{figure}[tbp]
\centering
\includegraphics[height=10cm,width=8.5cm]{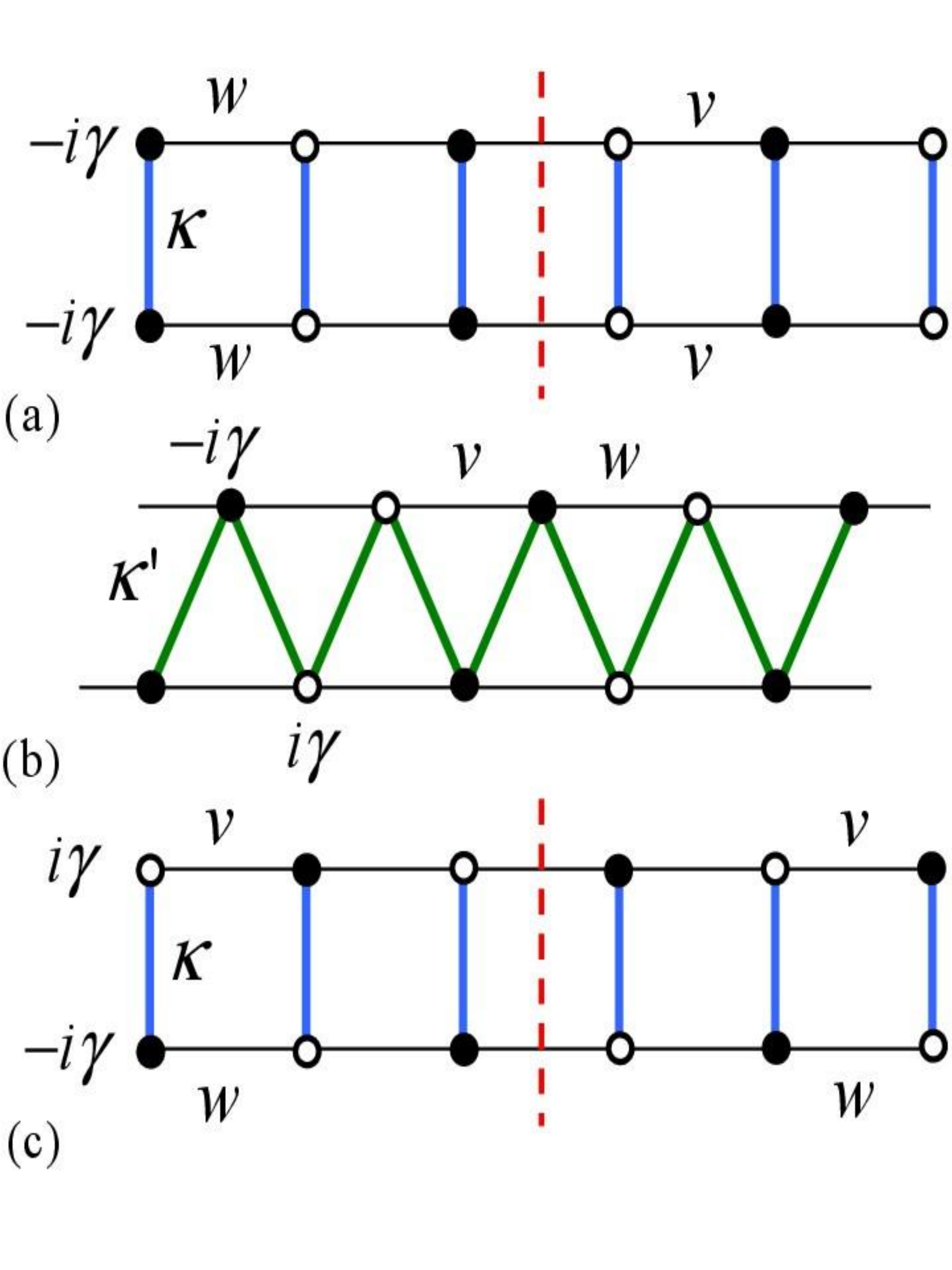}
\caption{(Color online) Schematic illustration of three types of regular
ladder systems. (a) and (c) have PT symmetry, while (b) is irrelevant to the
symmetry involving the bilinear operator T. The red dot lines indicate the
axis of reflection operation. }
\label{fig2}
\end{figure}

%%%%%%%%%%%%%%%%%%%%%%%%%%%%%%%%%%%%%%%%%%%%%%%%%%%%%%%%%%%%%%%%%%%%%%%

\section{Dimerized phase}

\label{Dimerized phase}

In this section, we investigate another type of uniform ladders, which is
illustrated in Fig. 2(c). The Hamiltonian reads

\begin{eqnarray}
&&H_{\mathrm{D}}=\sum\limits_{l=1}^{N}v\left( \left\vert 2l-1,1\right\rangle
\left\langle 2l,1\right\vert +\left\vert 2l,2\right\rangle \left\langle
2l+1,2\right\vert \right)  \notag \\
&&+\sum\limits_{l=1}^{N}w\left( \left\vert 2l,1\right\rangle \left\langle
2l+1,1\right\vert +\left\vert 2l-1,2\right\rangle \left\langle
2l,2\right\vert \right)  \notag \\
&&+\kappa \sum\limits_{l=1}^{2N}\left\vert l,1\right\rangle \left\langle
l,2\right\vert +\mathrm{H.c.}  \notag \\
&&+i\gamma \sum\limits_{l=1}^{2N}(-1)^{l+\lambda }\left\vert l,\lambda
\right\rangle \left\langle l,\lambda \right\vert \mathrm{,}
\end{eqnarray}%
where the boundary condition is $\left\vert 2N+1,\lambda \right\rangle
\equiv \left\vert 1,\lambda \right\rangle $. As illustrated in Fig. 2(c),
there is still a $\mathcal{PT}$-symmetry defined as before. According to the
non-Hermitian quantum theory, such a Hamiltonian may have fully real
spectrum within a certain parameter region. The boundary of the region is
the critical point of quantum phase transition associated with $\mathcal{PT}$%
-symmetry breaking. In the following, we will diagonalize this Hamiltonian
and investigate the phase diagram.

%%%%%%%%%%%%%%%%%%%%%%%%%%%%%%%%%%%%%%%%%%%%%%%%%%%%%%%%%%%%%%%%%%%%%%%
\begin{figure*}[tbp]
\centering
\includegraphics[height=6cm,width=5cm]{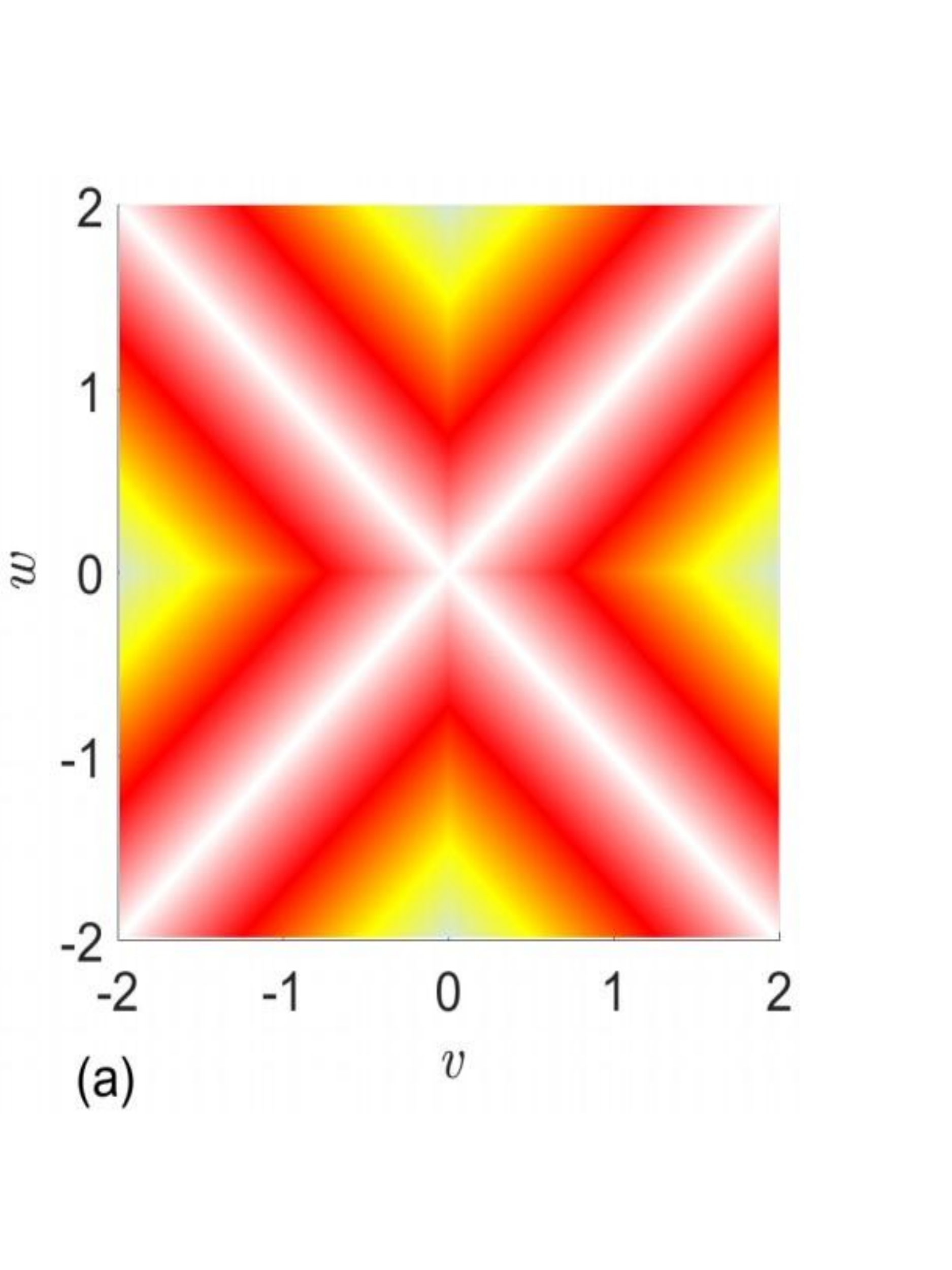}
\includegraphics[height=6cm,width=5cm]{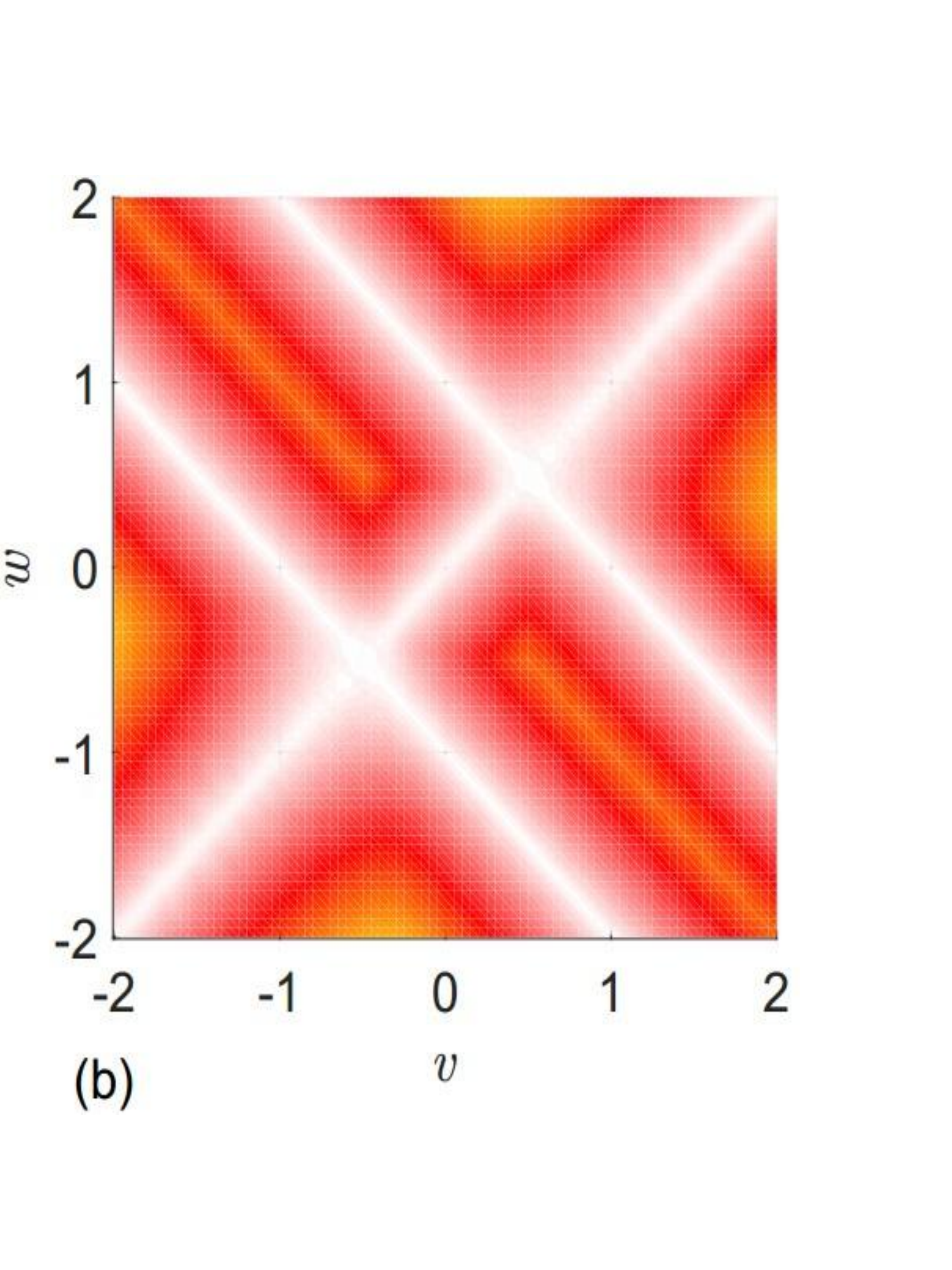}
\includegraphics[height=6cm,width=5cm]{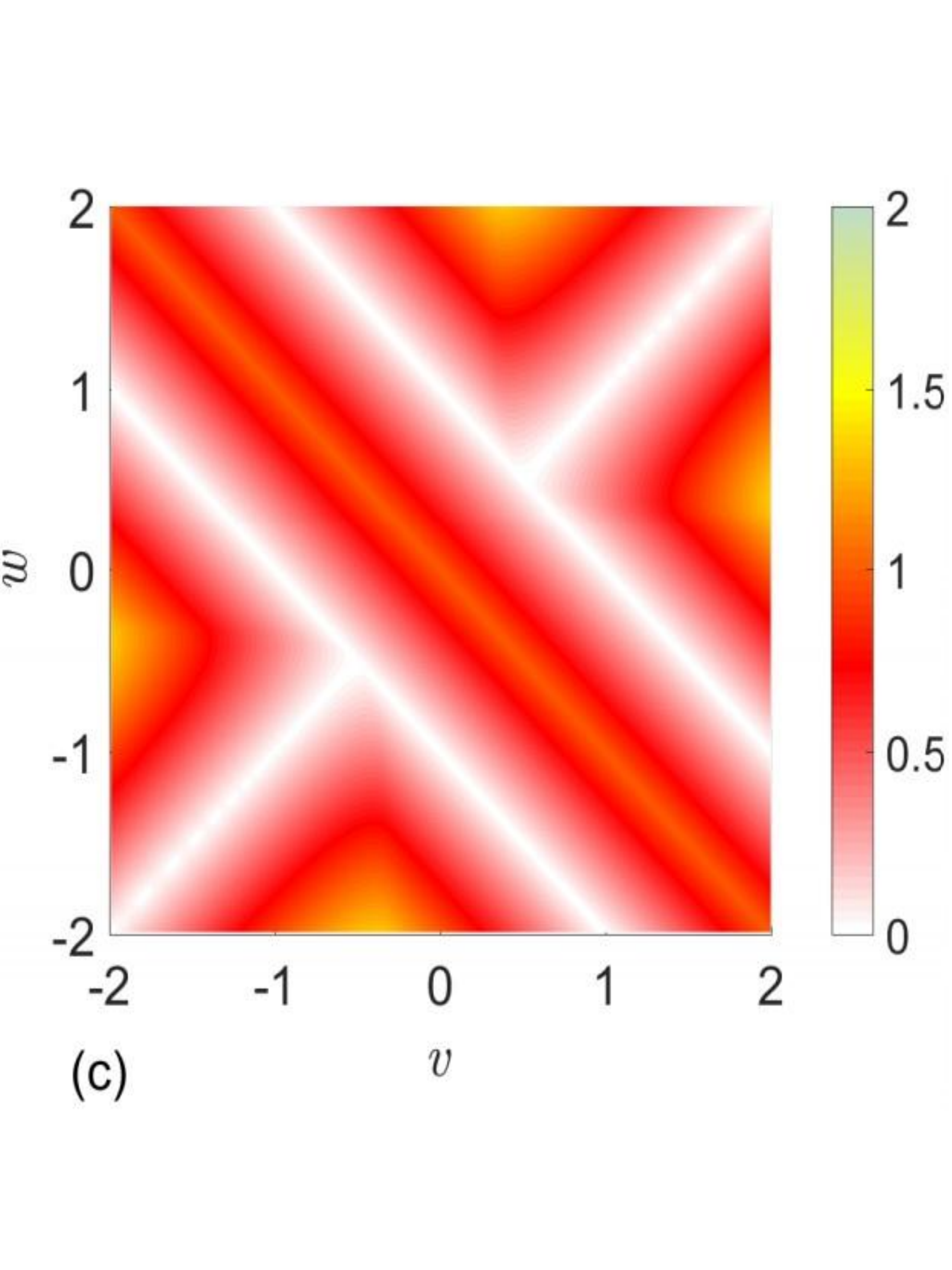}
\caption{(Color online) Phase diagrams of the systems $H_{\mathrm{T}}$, $H_{%
\mathrm{C}}$ and $H_{\mathrm{D}}$, which are presented by the distributions
of critical $\protect\gamma _{\mathrm{c}}$ on the $vw$ plane (in unit of $%
\protect\kappa $). Light regions correspond to small or zero $\protect\gamma %
_{\mathrm{c}}$, at which the reality of the spectrum is fragile. Beyond
these regions,\ there always exist common parameters $(w,v,\protect\gamma )$%
\ to maintain the full real spectrum for three types of non-Hermitian
systems. This fact ensures the probable existence of stable dynamical
signature of moire pattern.}
\label{fig3}
\end{figure*}
%%%%%%%%%%%%%%%%%%%%%%%%%%%%%%%%%%%%%%%%%%%%%%%%%%%%%%%%%%%%%%%%%%%%%%%
Taking the transformation

\begin{equation}
\left\{
\begin{array}{c}
\left\vert k,a\right\rangle =\sum\limits_{l=1}^{N}\frac{e^{i2kl}}{\sqrt{2N}}%
(e^{-ik}\left\vert 2l-1,1\right\rangle +\left\vert 2l,2\right\rangle ) \\
\left\vert k,b\right\rangle =\sum\limits_{l=1}^{N}\frac{e^{i2kl}}{\sqrt{2N}}%
(\left\vert 2l,1\right\rangle +e^{-ik}\left\vert 2l-1,2\right\rangle )%
\end{array}%
\right. ,
\end{equation}%
we have%
\begin{equation}
H_{\mathrm{D}}=\sum\limits_{k}(\left\vert k,a\right\rangle ,\left\vert
k,b\right\rangle )h_{k}\left(
\begin{array}{c}
\left\langle k,a\right\vert \\
\left\langle k,b\right\vert%
\end{array}%
\right) ,
\end{equation}%
where the kernel matrix is%
\begin{equation}
h_{k}=\left(
\begin{array}{cc}
i\gamma & we^{ik}+ve^{-ik}+\kappa \\
we^{-ik}+ve^{ik}+\kappa & -i\gamma%
\end{array}%
\right) .
\end{equation}%
The eigenvalue of $h_{k}$\ is
\begin{equation}
\varepsilon _{k}=\pm \sqrt{\left( \varepsilon _{k}^{0}\right) ^{2}-\gamma
^{2}},
\end{equation}%
where%
\begin{equation}
\varepsilon _{k}^{0}=\left\vert we^{-ik}+ve^{ik}+\kappa \right\vert ,
\end{equation}%
is eigenvalues of $h_{k}^{0}=h_{k}(\gamma =0)$. It has been shown that $%
\varepsilon _{k}^{0}$\ is always nonzero except at the lines%
\begin{equation}
w=v\text{, }\left\vert 2v/\kappa \right\vert >1,  \label{boundary1}
\end{equation}%
and%
\begin{equation}
\left\vert \left( w+v\right) /\kappa \right\vert =1.  \label{boundary2}
\end{equation}%
The phase diagram is plotted in Fig. \ref{fig3}(c) \cite{LC}. Then $%
\varepsilon _{k}$\ can be real within the whole $wv$ plane, once a
appropriate $\gamma $\ is taken. There is only one type of dimerization,
inter- dimer. Then it is referred as\ to dimerized phase. The extremely
dimerized phase is characterized by the ground states of the system with $%
\nu =w=0$ (see Fig. \ref{fig1}(d2)).

\section{Crossover phase}

\label{Crossover phase}In this section, we investigate the third type of
uniform ladders, which is a crossover between two types of structures above
and illustrated in Fig. 2(b). The Hamiltonian reads

\begin{eqnarray}
&&H_{\mathrm{C}}=\sum\limits_{l=1}^{N}\sum\limits_{\lambda =1,2}\left(
w\left\vert 2l-1,\lambda \right\rangle \left\langle 2l,\lambda \right\vert
+v\left\vert 2l,\lambda \right\rangle \left\langle 2l+1,\lambda \right\vert
\right)  \notag \\
&&+\kappa ^{\prime }\sum\limits_{l=1}^{2N}\left( \left\vert l,1\right\rangle
\left\langle l,2\right\vert +\left\vert l,1\right\rangle \left\langle
l+1,2\right\vert \right) +\mathrm{H.c.}  \notag \\
&&+i\gamma \sum\limits_{l=1}^{2N}\sum\limits_{\lambda
=1,2}(-1)^{l}\left\vert l,\lambda \right\rangle \left\langle l,\lambda
\right\vert \mathrm{.}
\end{eqnarray}%
Taking the Fourier transformation%
\begin{equation}
\left\{
\begin{array}{c}
\left\vert 2l-1,1\right\rangle =\frac{1}{\sqrt{N}}\sum_{k}e^{ikl}\left\vert
k,a\right\rangle \\
\left\vert 2l,1\right\rangle =\frac{1}{\sqrt{N}}\sum_{k}e^{ikl}\left\vert
k,b\right\rangle \\
\left\vert 2l-1,2\right\rangle =\frac{1}{\sqrt{N}}\sum_{k}e^{ikl}\left\vert
k,c\right\rangle \\
\left\vert 2l,2\right\rangle =\frac{1}{\sqrt{N}}\sum_{k}e^{ikl}\left\vert
k,d\right\rangle%
\end{array}%
\right. ,
\end{equation}%
we get

\begin{equation}
H_{\mathrm{C}}=\sum_{k}\left\vert \psi _{k}\right\rangle h_{k}\left\langle
\psi _{k}\right\vert ,
\end{equation}%
where the $4\times 4$ matrix
\begin{equation}
h_{k}=\left(
\begin{array}{cccc}
-i\gamma & \lambda _{k} & \kappa ^{\prime } & \kappa ^{\prime } \\
\lambda _{-k} & i\gamma & \kappa ^{\prime }e^{-ik} & \kappa ^{\prime } \\
\kappa ^{\prime } & \kappa ^{\prime }e^{ik} & -i\gamma & \lambda _{k} \\
\kappa ^{\prime } & \kappa ^{\prime } & \lambda _{-k} & i\gamma%
\end{array}%
\right) ,
\end{equation}%
with $\lambda _{k}=w+ve^{ik}$, and the vector $\left\vert \psi
_{k}\right\rangle =(\left\vert k,a\right\rangle ,\left\vert k,b\right\rangle
,\left\vert k,c\right\rangle ,\left\vert k,d\right\rangle )$. It is hard to
get the simplified analytical expression of the eigen values of the matrix.
However, we only concern the difference of the phase diagram of $H_{\mathrm{C%
}}$\ from that of the above two systems. First of all, when taking $\nu =w$,
$H_{\mathrm{C}}$\ reduces to a uniform chain with staggered imaginary
potentials. There is no energy gap in the spectrum of $H_{\mathrm{C}}$\ for $%
\gamma =0$. Then any nonzero $\gamma $\ can induce imaginary levels at $%
k=\pi $. The conclusion is true for all values of $\nu =w$, which is
different from the case of dimer.\ Secondly, when taking $\lambda
_{k}=w+ve^{ik}=\sigma \kappa ^{\prime }(\sigma =\pm )$, i.e., $k=0$\ and $%
w+v=\sigma \kappa ^{\prime }$, matrix $h_{k}$\ reduces to%
\begin{equation}
h^{\prime }=\kappa ^{\prime }\left(
\begin{array}{cccc}
-i\gamma ^{\prime } & \sigma 1 & 1 & 1 \\
\sigma 1 & i\gamma ^{\prime } & 1 & 1 \\
1 & 1 & -i\gamma ^{\prime } & \sigma 1 \\
1 & 1 & \sigma 1 & i\gamma ^{\prime }%
\end{array}%
\right) ,
\end{equation}%
with $\gamma ^{\prime }=\gamma /\kappa ^{\prime }$. It is easy to check that
two of four eigenvalues $\varepsilon _{\sigma }$\ \ can be expressed as

\begin{equation}
\varepsilon _{\sigma }=-\sigma 1\pm i\gamma^{\prime } ,
\end{equation}%
which indicate that any nonzero $\gamma $\ can induces complex levels . This
result is different from the case of tetramer. The phase diagram is plotted
in Fig. \ref{fig3}(b).

%FIG. 3. (color online) Phase diagrams of $H_{\mathrm{D}}$ and $H_{\mathrm{U}%
%} $ with zero $\gamma $ on the $vw$ plane (in unit of $\kappa $). Dot lines $%
%(w=\pm v)$ indicate the gapless point of $H_{\mathrm{D}}$\ at zero $\gamma $%
%. Solid lines indicate the zero point of $\varepsilon _{k}^{0}$ at which
%only zero $\gamma $\ permits the reality of $\varepsilon _{k}$. Beyond the
%lines $\varepsilon _{k}$,\ there always exists nonzero $\gamma $\ to maitain
%the full real spectrum for the non-Hermitian system.
%%%%%%%%%%%%%%%%%%%%%%%%%%%%%%%%%%%%%%%%%%%%%%%%%%%%%%%%%%%%%%%%%%%%%%%

%%%%%%%%%%%%%%%%%%%%%%%%%%%%%%%%%%%%%%%%%%%%%%%%%%%%%%%%%%%%%%%%%%%%%%%
%FIG. 4. Phase diagrams of the Hamiltonian in Eq. () with some representative
%$\gamma $ values. Panels (a)--(d) show . Point indicates the parameters
%under which the numerical simulation of time evolution in Fig. ? is
%performed.

\section{Dynamical signatures}

\label{Dynamical signatures}

In this section, we investigate the dynamics of the regular ladder systems
in the limit case $\kappa $, $\kappa ^{\prime }\gg w\gg \nu $. We start to
reduce the original lattices by decoupling between tetramers in $H_{\mathrm{T%
}}$\ by taking $v=0$\ and dimers in $H_{\mathrm{D}}$ by taking $v=w=0$. We
employ Hamiltonians $h_{\mathrm{T}}$\ and $h_{\mathrm{D}}$\ to describe two
clusters, which are schematically illustrated in Fig. \ref{fig1}(b2) and
(d2). In the following, we study the two sub-Hamiltonians.

\subsection{Tetramer cluster}

The $4$-site Hamiltonian $h_{\mathrm{T}}$\ reads

\begin{eqnarray}
h_{\mathrm{T}} &=&w\left( \left\vert 1,1\right\rangle \left\langle
2,1\right\vert +\left\vert 1,2\right\rangle \left\langle 2,2\right\vert
\right)  \notag \\
&&+\kappa (\left\vert 1,1\right\rangle \left\langle 1,2\right\vert
+\left\vert 2,1\right\rangle \left\langle 2,2\right\vert )+\mathrm{H.c.}
\notag \\
&&+i\gamma \sum\limits_{l=1}^{2}\sum\limits_{\lambda =1,2}(-1)^{l}\left\vert
l,\lambda \right\rangle \left\langle l,\lambda \right\vert \mathrm{,}
\label{h_t}
\end{eqnarray}%
or the matrix form%
\begin{equation}
h_{\mathrm{T}}=\left(
\begin{array}{cccc}
-i\gamma & \kappa & w & 0 \\
\kappa & -i\gamma & 0 & w \\
w & 0 & i\gamma & \kappa \\
0 & w & \kappa & i\gamma%
\end{array}%
\right) \mathrm{,}  \label{h_T matrix}
\end{equation}%
based on the basis set $\left\{ \left\vert 1,1\right\rangle ,\left\vert
1,2\right\rangle ,\left\vert 2,1\right\rangle ,\left\vert 2,2\right\rangle
\right\} $. The eigenvectors can be obtained explicitly as%
\begin{equation}
\left(
\begin{array}{c}
\left\vert \chi _{++}\right\rangle \\
\left\vert \chi _{+-}\right\rangle \\
\left\vert \chi _{-+}\right\rangle \\
\left\vert \chi _{--}\right\rangle%
\end{array}%
\right) =\left(
\begin{array}{cccc}
\Lambda ^{\ast } & \Lambda ^{\ast } & w & w \\
-\Lambda & -\Lambda & w & w \\
-\Lambda ^{\ast } & \Lambda ^{\ast } & -w & w \\
\Lambda & -\Lambda & -w & w%
\end{array}%
\right) \left(
\begin{array}{c}
\left\vert 1,1\right\rangle \\
\left\vert 1,2\right\rangle \\
\left\vert 2,1\right\rangle \\
\left\vert 2,2\right\rangle%
\end{array}%
\right) ,
\end{equation}%
with corresponding eigenvalues%
\begin{equation}
\varepsilon _{\sigma \rho }=\sigma \kappa +\rho \overline{w},(\sigma ,\rho
=\pm ),
\end{equation}%
where $\Lambda =\overline{w}+i\gamma $ and$\ \overline{w}=\sqrt{w^{2}-\gamma
^{2}}$. The dynamics of the system depends on eigen levels: For real $%
\overline{w}$, the average Dirac probability is conserved, while varies
exponentially for imaginary $\overline{w}$. The fascinating behavior occurs
at $\overline{w}=0$, i.e., $w=\gamma $, which is the EP of the system. We
can see that
\begin{equation}
\left\vert \chi _{++}\right\rangle =\left\vert \chi _{+-}\right\rangle
,\left\vert \chi _{-+}\right\rangle =\left\vert \chi _{--}\right\rangle
\mathrm{,}
\end{equation}%
from the expression of $\chi _{\sigma \rho }$\ when taking $\gamma =\gamma
_{c}=w$. Two pairs of eigenvectors coalescence to a single pair. It has been
shown that the Dirac probability increases quadratically with time \cite%
{WangP}.

\begin{figure*}[tbp]
\centering
\includegraphics[height=7cm,width=5cm]{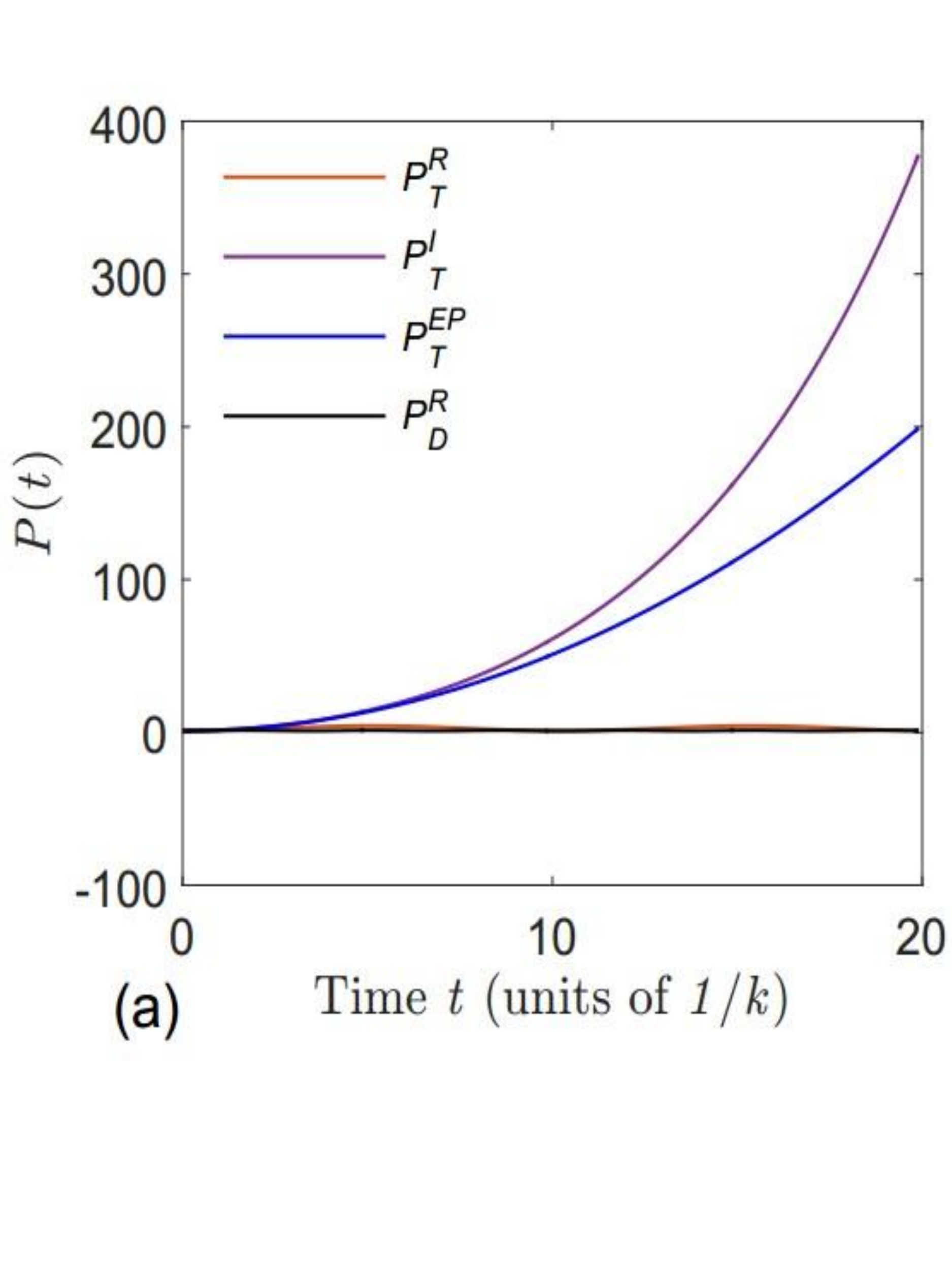}
\includegraphics[height=7cm,width=5cm]{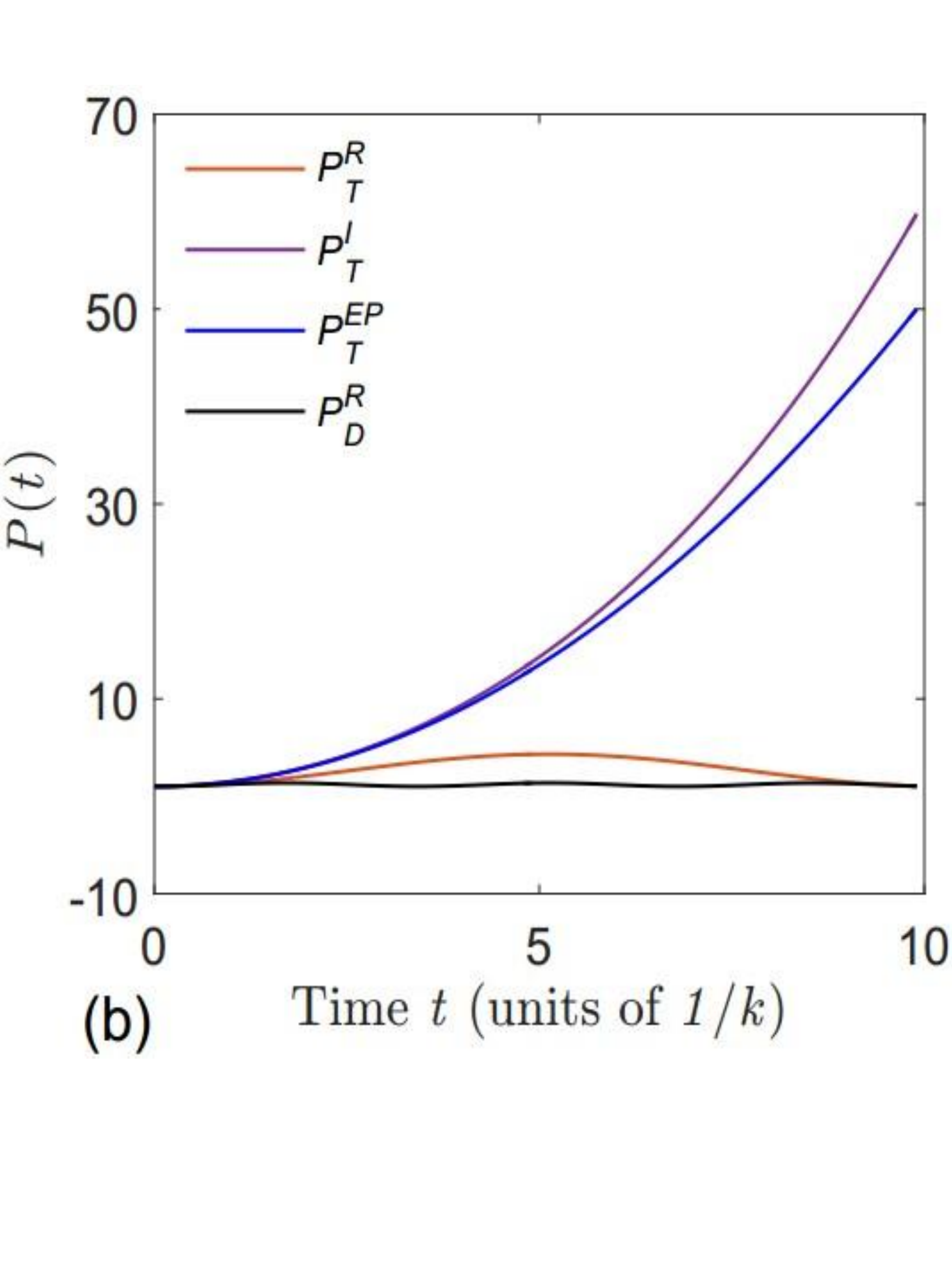}
\includegraphics[height=7cm,width=5cm]{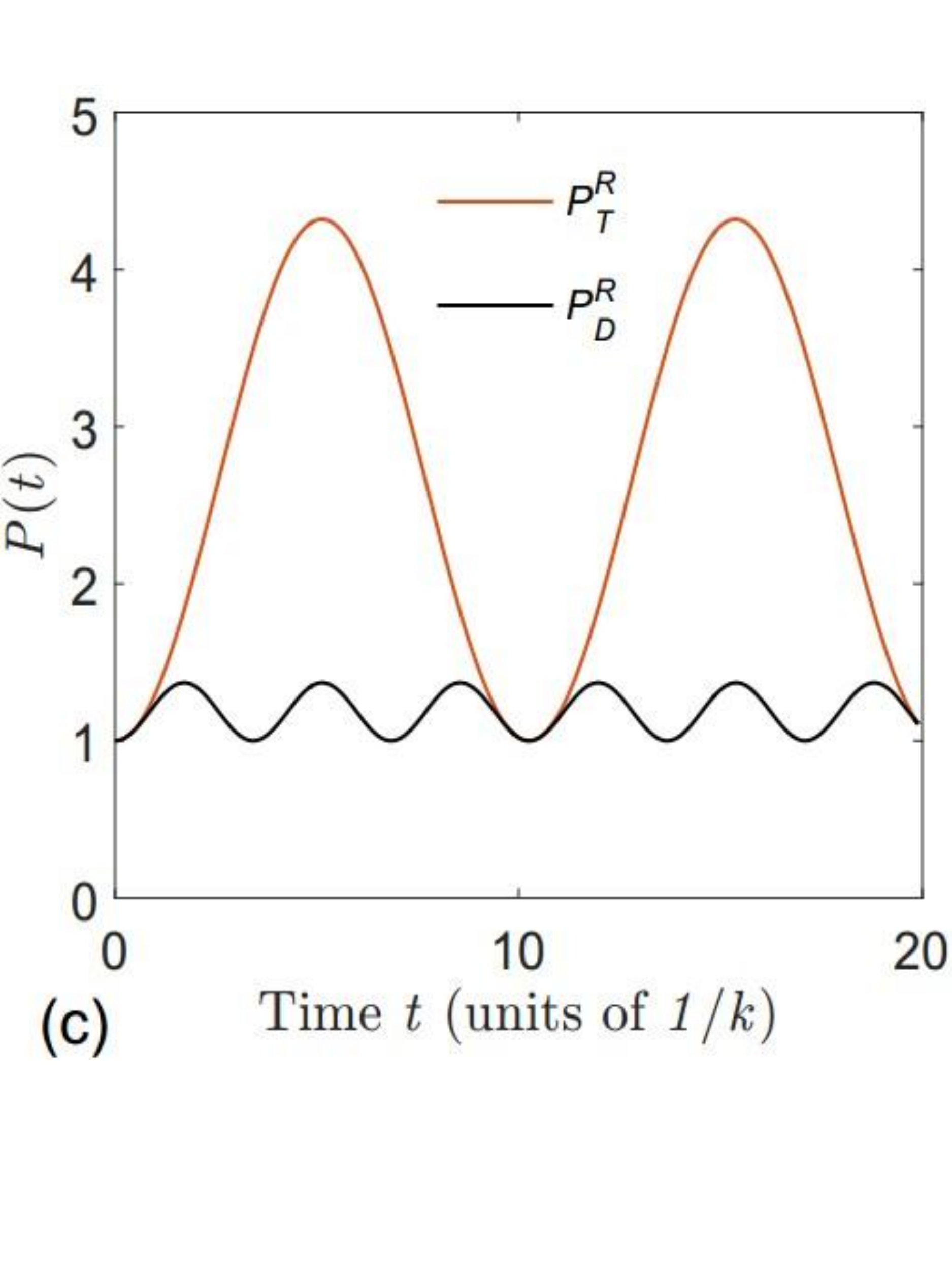}
\caption{(Color online) Profiles of Dirac probabilities of time evolution
for four typical dynamical behaviors. Plots of $P_{\mathrm{T}}^{\mathrm{R}%
}(t)$,\ $P_{\mathrm{T}}^{\mathrm{I}}(t)$, $P_{\mathrm{T}}^{\mathrm{EP}}(t)$
and $P_{\mathrm{D}}^{\mathrm{R}}(t)$\ are obtained from Eq. (\protect\ref%
{PTR}) with $w=0.5,\protect\gamma =0.395,$ Eq. (\protect\ref{PTI}) with $%
w=0.5,\protect\gamma =0.505,$ Eq. (\protect\ref{PTEP}) with $w=\protect%
\gamma =0.5$ and Eq. (\protect\ref{PDR}) with $\protect\gamma =0.395$,
respectively. We take $\protect\kappa =1$\ for all cases. Panels (a)--(c)
show the same plots in different scales. We can see the four types of
systems exhibit distinguishable dynamical behaviors, which are building
blocks of moire pattern in Fig. \protect\ref{fig5}.}
\label{fig4}
\end{figure*}

To characterize the dynamics, we consider the time evolution of the initial
state%
\begin{equation}
\left\vert \psi (0)\right\rangle =\frac{1}{2}\left( \left\vert
1,1\right\rangle +\left\vert 1,2\right\rangle +\left\vert 2,1\right\rangle
+\left\vert 2,2\right\rangle \right) .  \label{initial state}
\end{equation}%
(i) For the real $\overline{w}$, the evolved state at instant $t$\ is%
\begin{eqnarray}
\left\vert \psi _{\mathrm{T}}(t)\right\rangle &=&\frac{e^{-i\kappa t}}{2}%
\{[\cos \left( \overline{w}t\right) -i\frac{w-i\gamma }{\overline{w}}\sin
\left( \overline{w}t\right) ](\left\vert 1,1\right\rangle  \notag \\
&&+\left\vert 1,2\right\rangle )+[\cos \left( \overline{w}t\right) -i\frac{%
w+i\gamma }{\overline{w}}\sin \left( \overline{w}t\right) ]  \notag \\
&&\left( \left\vert 2,1\right\rangle +\left\vert 2,2\right\rangle \right) \}.
\end{eqnarray}%
The Dirac probability is%
\begin{eqnarray}
P_{\mathrm{T}}^{\mathrm{R}}(t) &=&\left\vert \left\vert \psi _{\mathrm{T}%
}(t)\right\rangle \right\vert ^{2}\   \notag \\
&=&\left( \frac{w}{\overline{w}}\right) ^{2}-\left( \frac{\gamma }{\overline{%
w}}\right) ^{2}\cos \left( 2\overline{w}t\right) ,  \label{PTR}
\end{eqnarray}%
which is a periodic function of time with the period $\tau _{\mathrm{T}}=\pi
/\overline{w}$.

(ii) For the imaginary $\overline{w}$, the evolved state at instant $t$\ is%
\begin{eqnarray}
\left\vert \psi _{\mathrm{T}}(t)\right\rangle &=&\frac{e^{-i\kappa t}}{2}%
\{[\cosh \left( \left\vert \overline{w}\right\vert t\right) -i\frac{%
w-i\gamma }{\left\vert \overline{w}\right\vert }\sinh \left( \left\vert
\overline{w}\right\vert t\right) ](\left\vert 1,1\right\rangle  \notag \\
&&+\left\vert 1,2\right\rangle )+[\cosh \left( \left\vert \overline{w}%
\right\vert t\right) -i\frac{w+i\gamma }{\left\vert \overline{w}\right\vert }%
\sinh \left( \left\vert \overline{w}\right\vert t\right) ]  \notag \\
&&\left( \left\vert 2,1\right\rangle +\left\vert 2,2\right\rangle \right) \}.
\end{eqnarray}%
The Dirac probability is%
\begin{eqnarray}
P_{\mathrm{T}}^{\mathrm{I}}(t) &=&\left\vert \left\vert \psi _{\mathrm{T}%
}(t)\right\rangle \right\vert ^{2}  \notag \\
&=&\left( \frac{\gamma }{\left\vert \overline{w}\right\vert }\right)
^{2}\cosh \left( 2\left\vert \overline{w}\right\vert t\right) -\left( \frac{w%
}{\left\vert \overline{w}\right\vert }\right) ^{2},  \label{PTI}
\end{eqnarray}%
which is an exponential function of time with the characteristic time
constant $\Gamma =1/\left\vert 2\overline{w}\right\vert $.

(iii) At the EP with zero $\overline{w}$, Jordan blocks appear in the matrix
$h_{\mathrm{T}}$. According to the appendix, the evolved state at instant $t$%
\ is%
\begin{eqnarray}
\left\vert \psi _{\mathrm{T}}(t)\right\rangle &=&\frac{e^{-i\kappa t}}{2}\{%
\left[ 1-t\gamma \left( 1+i\right) \right] \left( \left\vert
1,1\right\rangle +\left\vert 1,2\right\rangle \right)  \notag \\
&&+\left[ 1+t\gamma \left( 1-i\right) \right] \left( \left\vert
2,1\right\rangle +\left\vert 2,2\right\rangle \right) \}.
\end{eqnarray}%
The Dirac probability is%
\begin{equation}
P_{\mathrm{T}}^{\mathrm{EP}}(t)=\left\vert \left\vert \psi _{\mathrm{T}%
}(t)\right\rangle \right\vert ^{2}=1+2\gamma ^{2}t^{2},  \label{PTEP}
\end{equation}%
which increases quadratically with time, as the form $t^{2}$. Plots of three
typical dynamical behaviors are presented in Fig. \ref{fig4}.

\subsection{Dimer cluster}

The $2$-site Hamiltonian $h_{\mathrm{D}}$\ reads

\begin{eqnarray}
h_{\mathrm{D}} &=&\kappa \left\vert 1,1\right\rangle \left\langle
1,2\right\vert +\mathrm{H.c.}  \notag \\
&&+i\gamma \left( \left\vert 1,1\right\rangle \left\langle 1,1\right\vert
-\left\vert 1,2\right\rangle \left\langle 1,2\right\vert \right) .
\label{h_D}
\end{eqnarray}%
The eigenvectors can be obtained explicitly as%
\begin{equation}
\left(
\begin{array}{c}
\left\vert \chi _{+}\right\rangle \\
\left\vert \chi _{-}\right\rangle%
\end{array}%
\right) =\left(
\begin{array}{cc}
i\gamma +\varepsilon _{+} & \kappa \\
i\gamma +\varepsilon _{-} & \kappa%
\end{array}%
\right) \left(
\begin{array}{c}
\left\vert 1,1\right\rangle \\
\left\vert 1,2\right\rangle%
\end{array}%
\right) ,
\end{equation}%
with corresponding eigenvalues

\begin{equation}
\varepsilon _{\sigma }=\sigma \sqrt{\kappa ^{2}-\gamma ^{2}},(\sigma =\pm ).
\end{equation}%
In parallel, we consider the time evolution of the initial state

\begin{equation}
\left\vert \varphi _{\mathrm{D}}(0)\right\rangle =\frac{1}{\sqrt{2}}\left(
\left\vert 1,1\right\rangle +\left\vert 1,2\right\rangle \right) ,
\end{equation}%
which is a part of $\left\vert \psi (0)\right\rangle $. The evolved state is
\begin{eqnarray}
\left\vert \varphi _{\mathrm{D}}(t)\right\rangle &=&\frac{1}{\sqrt{2}}\{%
\left[ \cos \left( \varepsilon _{\mathrm{D}}t\right) -i\frac{\kappa +i\gamma
}{\varepsilon _{\mathrm{D}}}\sin \left( \varepsilon _{\mathrm{D}}t\right) %
\right] \left\vert 1,1\right\rangle  \notag \\
&&+\left[ \cos \left( \varepsilon _{\mathrm{D}}t\right) -i\frac{\kappa
-i\gamma }{\varepsilon _{\mathrm{D}}}\sin \left( \varepsilon _{\mathrm{D}%
}t\right) \right] \}\left\vert 1,2\right\rangle .
\end{eqnarray}%
The Dirac probability is%
\begin{eqnarray}
P_{\mathrm{D}}^{\mathrm{R}}(t) &=&\left\vert \left\vert \varphi _{\mathrm{D}%
}(t)\right\rangle \right\vert ^{2}  \notag \\
&=&\left( \frac{\kappa }{\varepsilon _{\mathrm{D}}}\right) ^{2}-\left( \frac{%
\gamma }{\varepsilon _{\mathrm{D}}}\right) ^{2}\cos \left( 2\varepsilon _{%
\mathrm{D}}t\right) ,  \label{PDR}
\end{eqnarray}%
which is a periodic function of time with the period $\tau _{\mathrm{D}}=\pi
/\varepsilon _{\mathrm{D}}$. In Fig. \ref{fig5} we plots the dynamical
behaviors in comparison with that of tetramer. The dynamics of such two
systems are distinct. In the parameter region $\kappa \gg w$, $\gamma \gg
\nu $, for real $\tau _{\mathrm{T}}$, we have $\tau _{\mathrm{T}}\gg \tau _{%
\mathrm{D}}$. And the tetramerized phase supports amplification of the
probability.

\section{Moire pattern}

\label{Moire pattern}

\begin{figure*}[tbp]
\includegraphics[height=10cm,width=8cm]{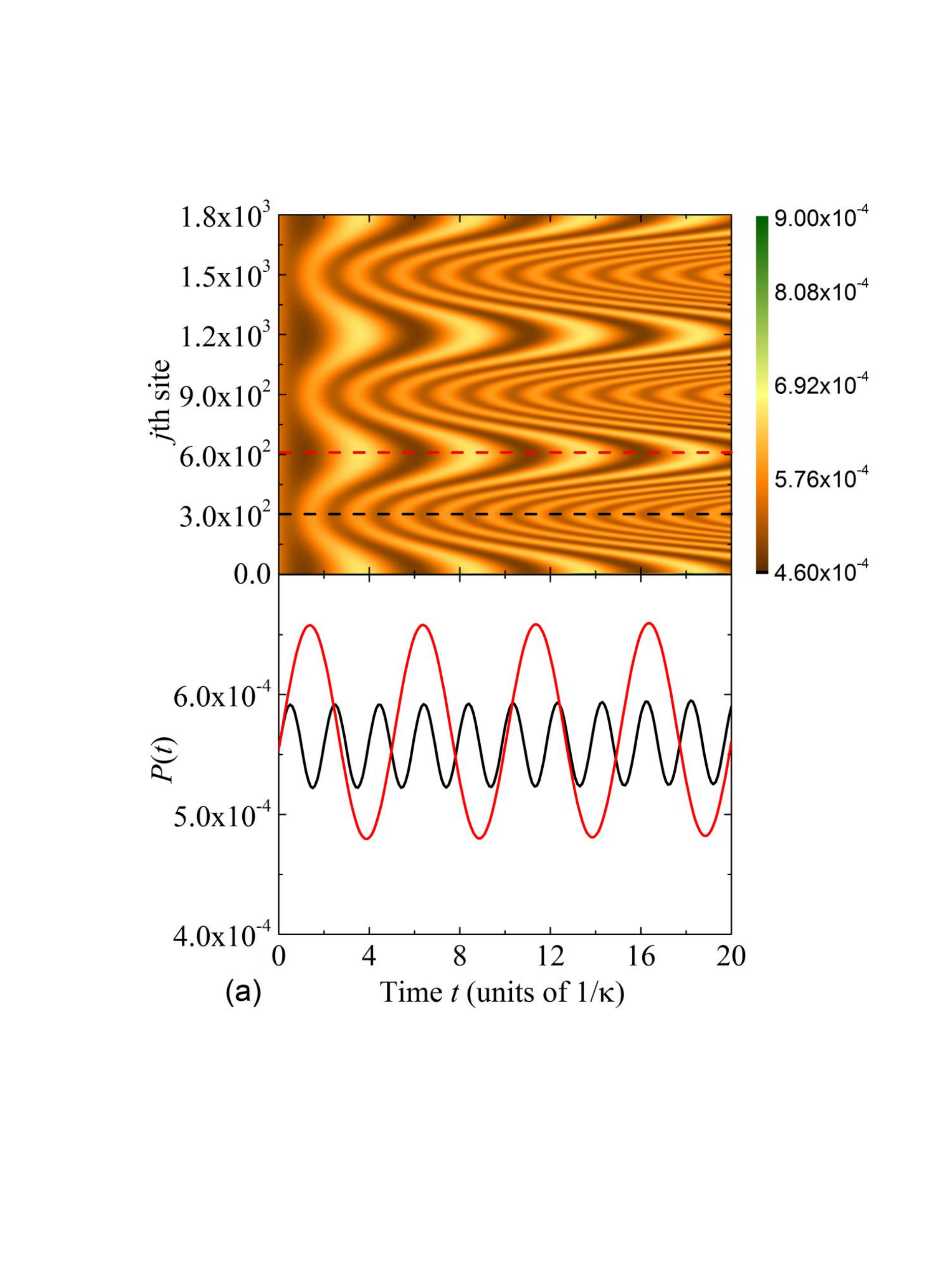}
\includegraphics[height=10cm,width=8cm]{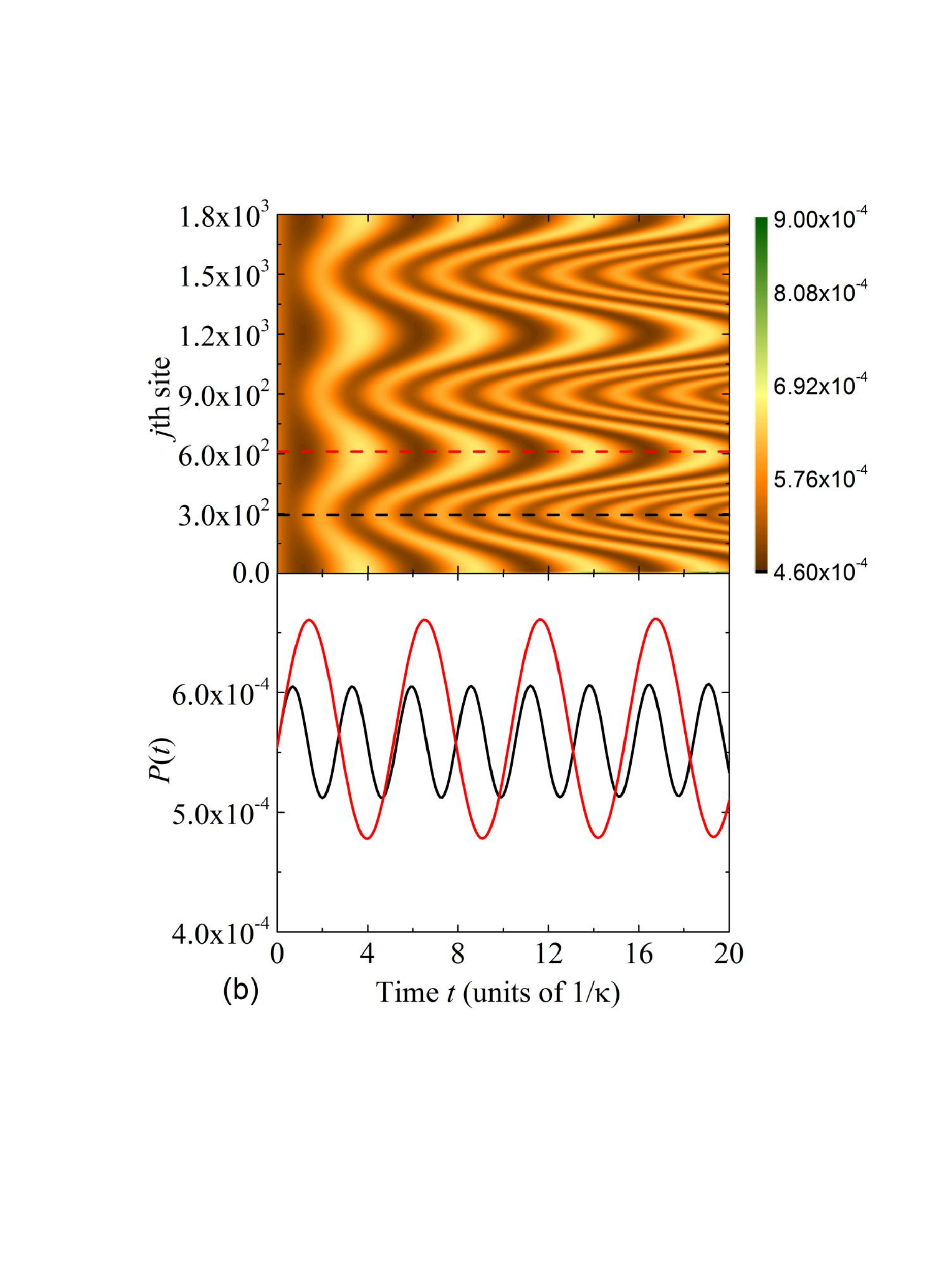}
\includegraphics[height=10cm,width=8cm]{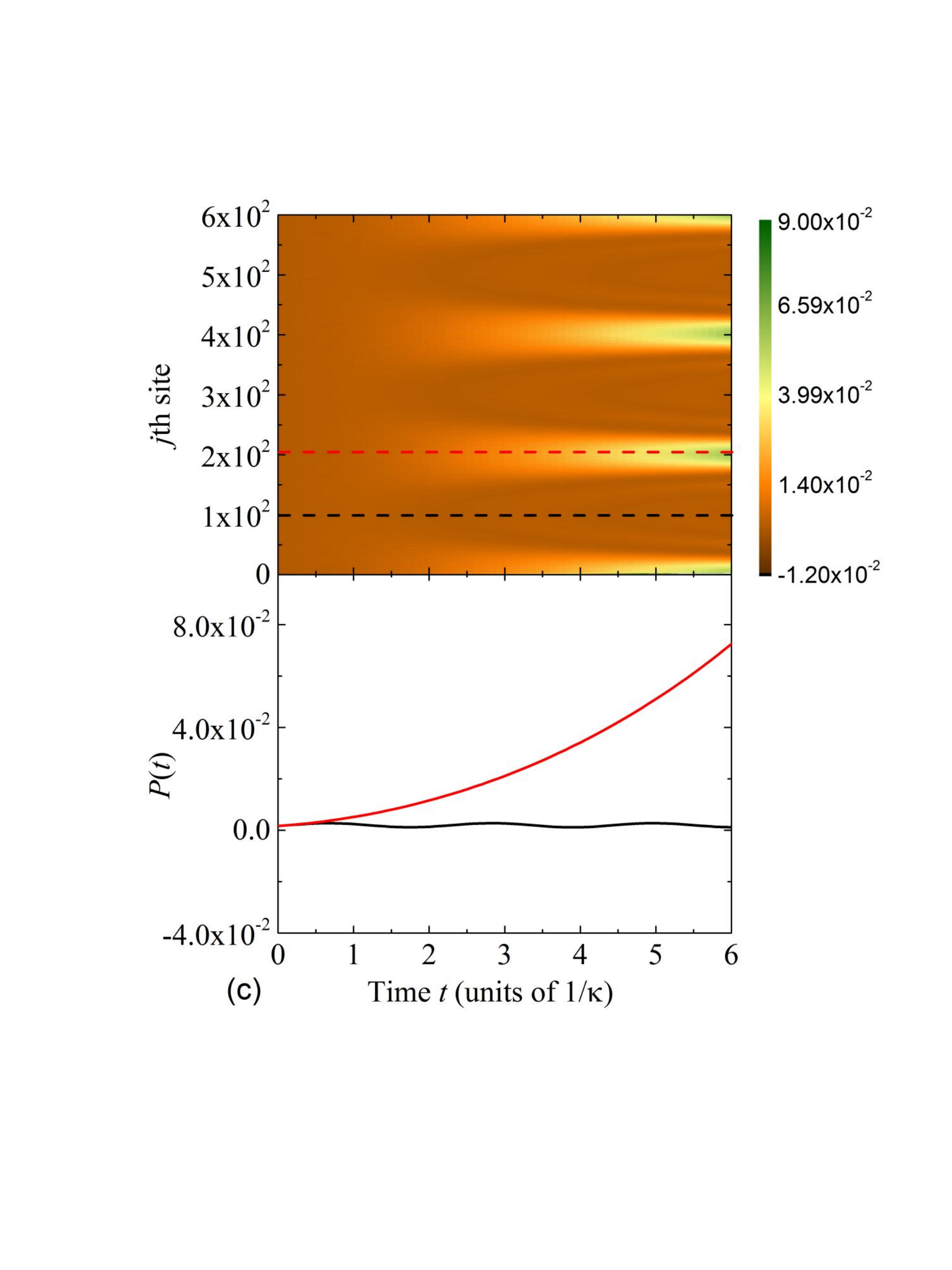}
\includegraphics[height=10cm,width=8cm]{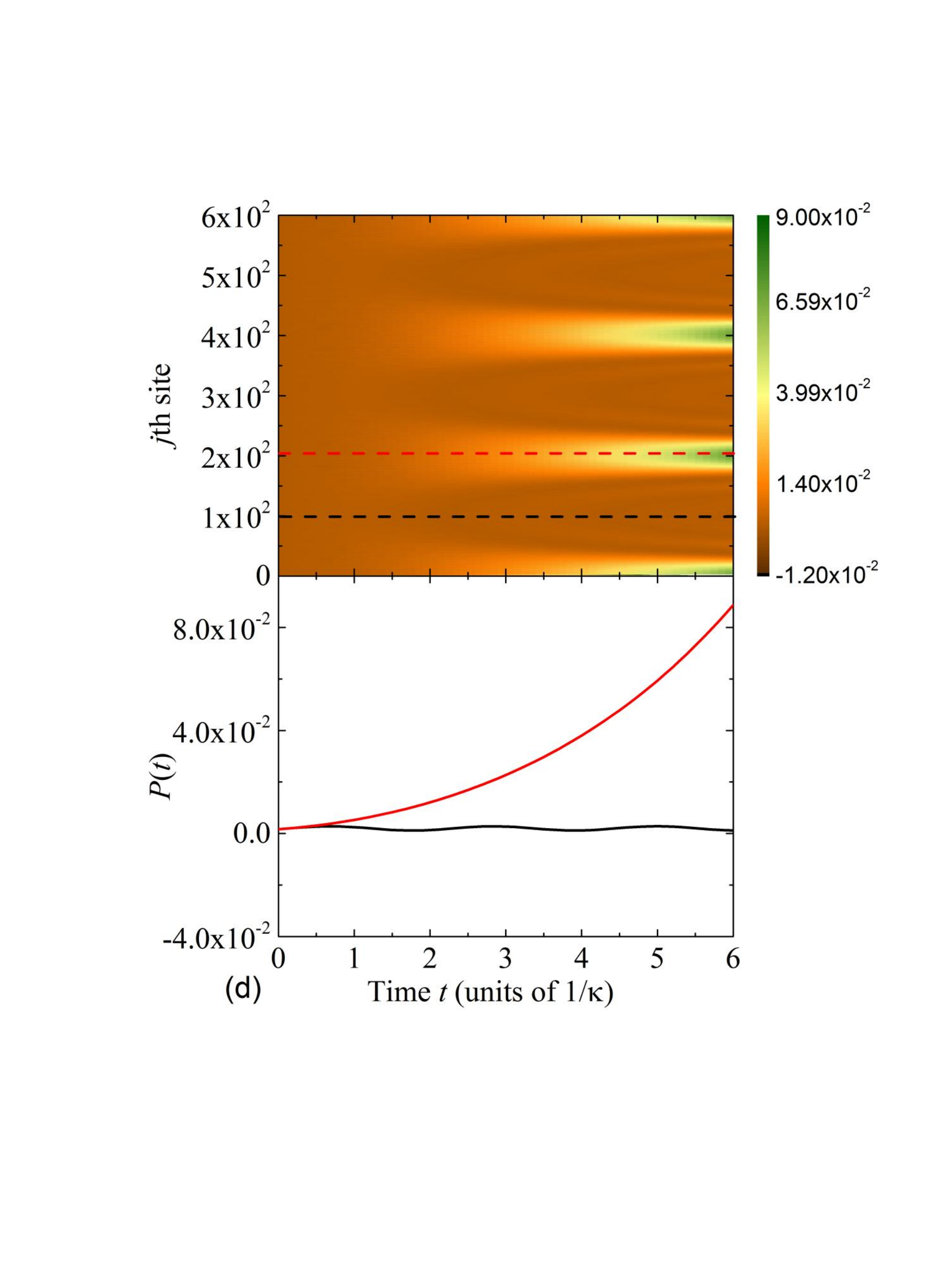}
\caption{(Color online) Propagation of the initial state being distributed
on each site with equal probability. The system parameters are (a) $\protect%
\kappa =1,$ $w=0.5,$ $v=0.1,$ $\protect\gamma =0.395,$ $\Delta =1/301$, (b) $%
\protect\kappa =0.6,$ $w=0.5,$ $v=0.1,$ $\protect\gamma =0.395,$ $\Delta
=1/301$, (c) $\protect\kappa =1,$ $w=0.5,$ $v=0.1,$ $\protect\gamma =0.645,$
$\Delta =1/101$, and (d) $\protect\kappa =1,$ $w=0.5,$ $v=0.1,$ $\protect%
\gamma =0.655,$ $\Delta =1/101$ respectively. The dashed red and black lines
in each panel denote the two distinct dynamical behaviors which are
determined by the tetramerized and dimerized substructures. The panels (a)
and (b) show that probability $P\left( j,t\right) $ varied periodically with
different period as time $t$ increases. In panels (c)-(d), one can see that
the Dirac probability increases in the tetramerized phase while it exhibits
periodical oscillation in the dimerized phase. Note that $P\left( t\right) $%
\ in the tetramerized phase increases quadratically with time $t$ in panel
(c), which is associated with the EP as given in Eq. (\protect\ref{PTEP}).
In comparision, $P\left( t\right) $ increases exponentially in the broken
phase of the tetramerized region, which is in agreement with the Eq. (%
\protect\ref{PTI}). These four panels show apparently that the slight
mismatched lattice constants lead to the long period moire patterns.}
\label{fig5}
\end{figure*}

In the original system, the inter-leg hopping rates are $\kappa $, $\kappa
^{\prime }$, $\kappa _{1}$ and $\kappa _{2}$, in various regions,
respectively. Three types of regular lattices emerge periodically along the
legs. Based on the above analysis, it is expected that the dynamical
behaviors in two cases are still distinct for nonzero $w$ and $v$. To
demonstrate this point, we perform a numerical simulation for the dynamics
of the original Hamiltonian. We take the initial state being distributed on
each site with the equal probability%
\begin{equation}
\left\vert \psi (0)\right\rangle =\frac{1}{\sqrt{2N}}\sum\limits_{l}^{N}%
\left( \left\vert l,1\right\rangle +\left\vert l,2\right\rangle \right) .
\end{equation}%
The evolved state is%
\begin{equation}
\left\vert \psi (t)\right\rangle =e^{-iHt}\left\vert \psi (0)\right\rangle ,
\end{equation}%
based on which the average value of all physical quantities can be obtained.
Our primary interest here is the influence of the moire pattern on the
dynamics of the system, or the mapping between the periodic structure and
the time-dependent Dirac probability distribution%
\begin{equation}
P(j,t)=\sum\limits_{\lambda =1,2}\left\vert \left\langle j,\lambda
\right\vert \psi (t)\rangle \right\vert ^{2}.
\end{equation}%
The inter-chain hopping rate is taken as the form
\begin{equation}
\kappa _{ij}=\kappa _{0}e^{-\alpha ^{2}[x_{i}-y_{j}]^{2}}=\kappa
_{0}e^{-\alpha ^{2}[i-(1-\Delta )j]^{2}},
\end{equation}%
where $x_{i}=i$, $y_{j}=(1-\Delta )j$ denote the dimensionless lattice site
coordinates in chain $1$ and $2$, respectively. Here $\alpha $ controls the
range of inter-chain tunneling amplitude $\kappa _{ij}$. We take $\alpha =2$%
, which\ results in $\kappa =\kappa _{0},\kappa ^{\prime }=0.37\kappa _{0}$
and ensures $\kappa _{ij}\approx 0$\ except the rates $\kappa $, $\kappa _{1}
$, $\kappa _{2}$, and $\kappa ^{\prime }$. The numerical simulation is
performed by exact diagonalization for a finite system with several typical
imaginary potentials $\gamma >\gamma _{c}$, $\gamma \approx \gamma _{c}$ and
$\gamma <\gamma _{c}$. The probability distributions $P(j,t)$ are plotted in
Fig. \ref{fig5}, which show that Moire patterns are apparent for each cases.
It is worth mentioning briefly that the moire pattern above discussed can be
implemented through the two dimensional array of evanescently coupled
optical waveguides with alternating regions of optical gain and absorption
\cite{K. G. Makris,Z. H. Musslimani,S. Klaiman,S. Longhi,Ganainy}. In this
context, the time evolution of state $\left\vert \psi (t)\right\rangle $
mapped into the spatial evolution of the modal amplitudes of light waves,
along the array axis. The hopping amplitude $w$,$\ v$, and $\kappa $ can be
modulated through inhomogeneous waveguide spacings.

\section{Summary}

\label{summary}

We have shown that a super periodicity in the coordinate space (along the
ladder) is imposed on two coupled SSH chains if there exists a slight
difference of lattice constants between two legs, in spite of a staggered
imaginary potential. There are two main gapped phases in each period, which
have full real spectra for appropriate system parameters. We have
characterized such phases within unbroken $\mathcal{PT}$-symmetry regions in
terms of the dimerization and tetramerization. Analytical analysis and
numerical simulation for the dynamics of two phases have shown that
imaginary potentials can enhance their distinguishability extremely. Hence,
the dynamics of the whole system is profoundly changed by slightly
mismatched lattice constants associated with long period moire patterns. Our
result for this concrete system proves insightful information about the
moire pattern in the non-Hermitian regime. In contrast to a Hermitian
system, the amplification of Dirac probability and parameter sensitivity of
a non-Hermitian system in the vicinity of EP demonstrate the remarkable
dynamic signature of moire patterns. The generalization to higher dimensions
is straight forward.

\section{Appendix}

\label{Appendix}

In this appendix we present the derivation of time evolution of the initial
state in Eq. (\ref{initial state}) under the system $h_{\mathrm{T}}$ (\ref%
{h_T matrix}) at the EP. It can be reduced to the problem of $2\times 2$
matrix%
\begin{equation}
M=\left(
\begin{array}{cc}
\kappa -i\gamma & w \\
w & \kappa +i\gamma%
\end{array}%
\right) ,
\end{equation}%
at $w=\gamma $. We have%
\begin{equation}
M=\left(
\begin{array}{cc}
\kappa -i\gamma & \gamma \\
\gamma & \kappa +i\gamma%
\end{array}%
\right) ,
\end{equation}%
which is in the Jordan block form%
\begin{eqnarray}
M &=&VhV^{-1}  \notag \\
&=&\left(
\begin{array}{cc}
-i\gamma & 1 \\
\gamma & 0%
\end{array}%
\right) \left(
\begin{array}{cc}
\kappa & 1 \\
0 & \kappa%
\end{array}%
\right) \left(
\begin{array}{cc}
0 & 1/\gamma \\
1 & i%
\end{array}%
\right) ,  \notag
\end{eqnarray}%
with%
\begin{equation*}
h=\left(
\begin{array}{cc}
\kappa & 1 \\
0 & \kappa%
\end{array}%
\right) ,V=\left(
\begin{array}{cc}
-i\gamma & 1 \\
\gamma & 0%
\end{array}%
\right) .
\end{equation*}%
The evolved state $\left\vert \Psi \left( t\right) \right\rangle $\ obeys
the Schrodinger equation%
\begin{equation}
i\frac{\mathrm{d}}{\mathrm{d}t}\left\vert \Psi \left( t\right) \right\rangle
=M\left\vert \Psi \left( t\right) \right\rangle ,
\end{equation}%
which leads to%
\begin{equation}
i\frac{\mathrm{d}}{\mathrm{d}t}\left\vert \Psi \left( t\right) \right\rangle
=VhV^{-1}\left\vert \Psi \left( t\right) \right\rangle ,
\end{equation}%
or%
\begin{equation}
i\frac{\mathrm{d}}{\mathrm{d}t}V^{-1}\left\vert \Psi \left( t\right)
\right\rangle =hV^{-1}\left\vert \Psi \left( t\right) \right\rangle .
\end{equation}%
Setting
\begin{equation*}
\left\vert \psi \right\rangle =\left(
\begin{array}{c}
\psi _{1} \\
\psi _{2}%
\end{array}%
\right) =V^{-1}\left\vert \Psi \left( t\right) \right\rangle ,
\end{equation*}%
we have%
\begin{equation}
i\frac{\mathrm{d}}{\mathrm{d}t}\left\vert \psi \right\rangle =h\left\vert
\psi \right\rangle ,
\end{equation}%
or the explicit form%
\begin{equation}
\left\{
\begin{array}{c}
i\frac{\mathrm{d}\psi _{1}}{\mathrm{d}t}=\kappa \psi _{1}+\psi _{2} \\
i\frac{\mathrm{d}\psi _{2}}{\mathrm{d}t}=\kappa \psi _{2}%
\end{array}%
\right. .
\end{equation}%
Then the solution is%
\begin{equation}
\left\{
\begin{array}{c}
\psi _{2}=c_{2}e^{-i\kappa t} \\
i\frac{\mathrm{d}\psi _{1}}{\mathrm{d}t}=\kappa \psi _{1}+c_{2}e^{-i\kappa t}
\\
\psi _{1}=c_{1}e^{-i\kappa t}-itc_{2}e^{-i\kappa t}%
\end{array}%
\right. ,
\end{equation}%
where $c_{1}$ and $c_{2}$ are constants determined by initial state. Finally
we have%
\begin{equation}
\left\vert \Psi \left( t\right) \right\rangle =V\left\vert \psi \left(
t\right) \right\rangle =V\left(
\begin{array}{c}
c_{1}e^{-i\kappa t}-itc_{2}e^{-i\kappa t} \\
c_{2}e^{-i\kappa t}%
\end{array}%
\right) .
\end{equation}%
For initial state
\begin{equation}
\left\vert \Psi \left( 0\right) \right\rangle =\left(
\begin{array}{c}
a \\
b%
\end{array}%
\right) ,
\end{equation}%
we have%
\begin{equation}
\left\vert \Psi \left( t\right) \right\rangle =e^{-i\kappa t}\left(
\begin{array}{c}
\left( 1-t\gamma \right) a-it\gamma b \\
\left( 1+t\gamma \right) b-it\gamma a%
\end{array}%
\right) .
\end{equation}%
Taking $a=b=1/\sqrt{2}$, we obtain%
\begin{equation}
\left\vert \Psi \left( t\right) \right\rangle =\frac{e^{-i\kappa t}}{\sqrt{2}%
}\left(
\begin{array}{c}
1-t\gamma \left( 1+i\right) \\
1+t\gamma \left( 1-i\right)%
\end{array}%
\right) .
\end{equation}%
It shows that the dynamics at EP is peculiar, which is linearly time
dependent.

\acknowledgments This work was supported by National Natural Science
Foundation of China (under Grants No. 11374163, No. 11505126). X.Z.Z. was
also supported by the Ph.D. research startup foundation of Tianjin Normal
University under Grant No. 52XB1415, and the Program for Innovative Research
in University of Tianjin (under Grant No. TD13-5077).

\end{document}